# Updatable Learned Indexes Meet Disk-Resident DBMS - From Evaluations to Design Choices

HAI LAN, RMIT University, Australia
ZHIFENG BAO*, RMIT University, Australia
J. SHANE CULPEPPER, RMIT University, Australia
RENATA BOROVICA-GAJIC, The University of Melbourne, Australia

Although many updatable learned indexes have been proposed in recent years, *whether they can outperform traditional approaches on disk remains unknown*. In this study, we revisit and implement four state-of-the-art updatable learned indexes on disk, and compare them against the B+-tree under a wide range of settings. Through our evaluation, we make some key observations: 1) Overall, the B+-tree performs well across a range of workload types and datasets. 2) A learned index could outperform B+-tree or other learned indexes on disk for a specific workload. For example, PGM achieves the best performance in write-only workloads while LIPP significantly outperforms others in lookup-only workloads. We further conduct a detailed performance analysis to reveal the strengths and weaknesses of these learned indexes on disk. Moreover, we summarize the observed common shortcomings in five categories and propose four design principles to guide future design of on-disk, updatable learned indexes: (1) reducing the index's tree height, (2) better data structures to lower operation overheads, (3) improving the efficiency of scan operations, and (4) more efficient storage layout.

## 1 INTRODUCTION

Driven by the promising in-memory performance profiles demonstrated in a pioneer work on learned index [12], several in-memory, updatable learned indexes [7, 9, 10, 30] have been proposed subsequently. Their in-memory superiority has also been verified in a recent comprehensive evaluation [29].

Meanwhile, many widely used Database Management Systems (DBMSs) still rely on disk-based operations for two main reasons: (1) the size of main memory is limited and the total size of the indexes can exceed the total RAM available on commodity hardware [4]; (2) in a DBMS, main memory is also used to perform expensive data processing operations, such as joins or sorting, or to perform data modeling and analysis. Thus, if all of the index data structures are loaded into

---

*corresponding author

Authors' addresses: Hai Lan, RMIT University, Melbourne, Victoria, Australia; Zhifeng Bao, RMIT University, Melbourne, Victoria, Australia; J. Shane Culpepper, RMIT University, Melbourne, Victoria, Australia; Renata Borovica-Gajic, The University of Melbourne, Melbourne, Victoria, Australia.

main memory as well, these operations can perform poorly, or in the worst case, fail to run at all. Moreover, the updatable, on-disk indexes are critical to support two common workload types – online transaction processing (OLTP) workloads and hybrid transaction/analytical processing (HTAP) workloads.

To this end, an important question still remains – *Can updatable learned indexes fully replace traditional on-disk indexes, such as the B+-tree?* However, there has not been any study trying to implement those updatable learned indexes on disk, not to mention conducting a comprehensive evaluation of their on-disk performance [13].

To answer this question, we for the first time study, extend, implement, and evaluate updatable learned indexes using one-dimensional data, on-disk, with two key goals: (1) To better understand how in-memory learned indexes perform in an on-disk setting; (2) To provide practitioners and researchers design decisions we have discovered when adapting learned indexes to disk-resident settings. In summary, this work makes the following contributions:

① **We compare and contrast four state-of-the-art, in-memory, updatable learned indexes, FITing-tree [10], ALEX [7], PGM [9], and LIPP [30]), and show how to extend and implement each of them as on-disk indexes.** Specifically, we first discuss how each index supports in-memory operations in Section 2. Then, in Section 3, we show how design decisions made in these indexes affect different types of workloads, such as data partitioning or searching in leaf nodes. Next, in Section 4, we show how to extend and implement each of them on disk, and discuss the External Memory (EM) model performance bounds for each learned index. Some of the learned indexes are difficult to re-implement as an on-disk data structure, e.g., ALEX being the most difficult.

② **We examine how on-disk operations affect learned indexes and compare each of them to the B+-Tree – one of the most efficient and commonly used on-disk data structures in the database community.** We perform a comprehensive evaluation in Sections 5-6, and test the indexes using eleven different datasets and six workload types – lookups, scan, inserts, heavy reads, heavy writes, and balanced reads and writes on two disk types, an HDD disk and an SDD disk. Those datasets exhibit a variety of different properties in terms of difficulty of being modeled with a linear function, which is the most common approach in most learned indexing models. We also study the impact of caching inner nodes of these indexes in main memory. Finally, we study the storage usage of each index, the impact of block size, and the robustness of each index by reporting the tail latencies. Based on our evaluation results, we create a set of observations (**O**), and provide key take-aways, which can be summarized as follows (detailed observations can be found in Section 6):

**K1. Performance on Disk (O1-O13) :** Each learned index has both strengths and weaknesses – yet none of them are competitive with the B+-tree across all tested workloads, in an on-disk setting.

**K2. Main Memory Impact (O14-O15):** When caching inner nodes in main memory, B+-tree can outperform learned indexes in all tested workloads, on all datasets, since the main overhead of learned indexes for search and insert is in "the last mile" – the leaf node traversal, which is consistently larger than a B+-tree leaf node.

**K3. Storage Usage (O16):** When including the leaf node size in a comparison, with the exception of PGM, existing learned indexes require more space than an on-disk B+-tree, and can even be 20x larger. The space used on disk cannot be reclaimed easily, and learned indexes require significant modifications to the index structure which further increases the storage size.

**K4. Impact of Block Size (O17) :** A larger block size can help reduce the total number of fetched blocks in the FITing-tree, ALEX, and PGM, but LIPP does not show any benefits from such a change.

**K5. Robustness (O18):** The B+-tree has more stable and a much smaller p99 latency in most evaluation settings, which translates to far fewer tail queries that are significantly slower than the average.

③ **We discuss important design decisions that must be considered when creating an on-disk, updatable learned index.** Based on our evaluation observations, we summarize the common shortcomings of on-disk, updatable learned indexes in Section 7.1. Finally, we propose four important design principles to guide the future design of on-disk, updatable learned indexes in Section 7.2:

(1) *reducing the index's height from root node to leaf node*;
(2) *better data structures to lower operation overheads*;
(3) *improving the efficiency of scan operations*;
(4) *more efficient storage layout*.

We believe these principles and our implementations [1] will help both researchers and practitioners to develop new disk-resident learned indexes which could be used in many practical applications.

## 2 UPDATABLE LEARNED INDEXES REVISITED

We now revisit four updatable learned indexes in detail and discuss other indexes related to our work. We broadly classify the them into two categories according to how they are constructed: **bottom-up**, which includes FITing-tree and PGM; and **top-down**, which includes ALEX and LIPP. Table 1 provides a taxonomy for them. For each index, we describe the index structure, followed by how lookup and insert operations are performed.

Table 1. A Taxonomy of Studied Indexes

| Index | Year | Inner Node | | Leaf Node | | Insert[2] | Data Partition | Node Size | Structure Modification | |
|---|---|---|---|---|---|---|---|---|---|---|
| | | Search Algo.[1] | Error | Search Algo. | Error | | | | Strategy | Memory Reuse |
| B+-tree | - | B.S | Tunable | B.S | Tunable | Empty Slot | Evenly | Tunable | Greedy | ✓ |
| FITing-tree [10] | 2019 | B.S | Tunable | Model + B.S | Tunable | Buffer | Greedy | Node-related | Greedy | × |
| PGM [9] | 2020 | Model + B.S | Tunable | Model + B.S | Tunable | Append/Rebuild | Streaming Algo. | N.A. | Greedy | × |
| ALEX [7] | 2020 | Model | Exact | Model + E.S | Unfixed | Gapped Array | - | Tunable | Cost-based | × |
| LIPP [30] | 2021 | Model[3] | Exact | Model | Exact | Gapped Array | - | Tunable[4] | Greedy | × |

[1] B.S, E.S, and Model denote binary search, exponential search, and predicting the position with a model, respectively.
[2] Here, we refer to how these indexes store the new insertion key-value pairs.
[3] LIPP does not distinguish the inner node and leaf node. We set the *Inner Node* and *Leaf Node* with same values.
[4] Although the authors state that a maximum node size is set, they could create a larger node size than the parameter (line 2 in Algorithm 5 [30]).

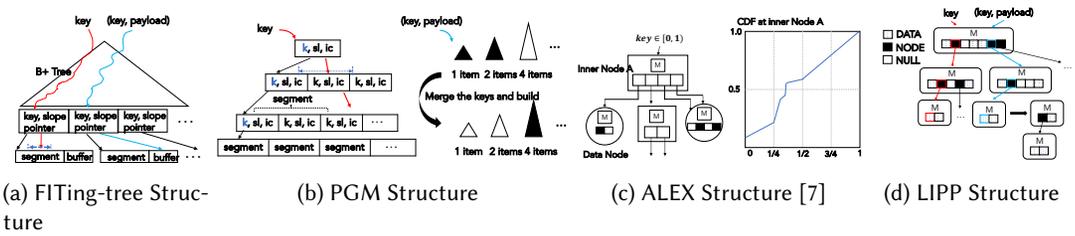

| (a) FITing-tree Structure | (b) PGM Structure | (c) ALEX Structure [7] | (d) LIPP Structure |

Fig. 1. Structures of the Learned Indexes Studied in This Work.

### 2.1 Learned Indexes Built Bottom-up

**FITing-tree** [10]. Figure 1(a) shows a FITing-tree's layout. Given an error bound, which indicates the maximum distance between the predicted position of one key and its true position, a FITing-tree first applies a greedy method to split the sorted array into segments, where each segment containing a linear model (the first key of the segment and the slope) to predict the position of a key. Then, a B+-tree is built over the keys covered by each linear model.

To support lookup operations, a FITing-tree first traverses from the root to a leaf node using a linear scan over inner nodes. After locating the segment that contains the search key, the learned linear model is used to predict the position *pred_pos*, and then a binary search is applied to the range [*pred_pos-error, pred_pos+error*].

To support insert operations, a FITing-tree has two strategies, an *Inplace Insert Strategy* and a *Delta Insert Strategy*. For the *Inplace Insert Strategy*, there are $\varepsilon$ empty slots added to the beginning and end of each segment. The index has a predefined error bound equal to $e + \varepsilon$, where $e$ is derived from the error from the linear model. If a segment is full, a greedy method is called on the key-payload pairs of the targeted segment to create a new segment using a *resegment* operation. The *Delta Insert Strategy* adds a buffer to every segment. New key-payload pairs are inserted into the buffer first. When the buffer is full, a *resegment* operation is triggered.

**PGM** [9]. Given a predefined error bound, PGM uses a streaming algorithm [23] rather than a greedy algorithm to create segments and the associated linear models (start key, slope, and intercept). Then, a streaming algorithm is applied recursively to construct the parent nodes for the keys in the models.

To support lookup operations, PGM first uses the model to predict a position *pos*, and then a binary search is applied on the range [*pred_pos-error, pred_pos+error*].

To support insert operations, a different insert mechanism is proposed for *Append-only Insert* and *Arbitrary Insert*. For an *Append-only Insert*, PGM first tries to add the new key into an end segment and checks if it is outside the targeted error bound. If not, the insert is complete; otherwise, a new segment is created with the new key and the parent nodes are updated accordingly. As shown in Figure 1(b), for an *Arbitrary Insert*, PGM maintains multiple indexes of different sizes simultaneously and adopts ideas from the LSM-tree to merge many small indexes into a larger one.

## 2.2 Learned Indexes Built Top-down

**ALEX** [7]. Figure 1(c) is an instance of an ALEX index as described in the original paper [7]. Alex has two node types – inner nodes and data nodes; both contain an array and a linear model to predict the positions in the associated array. The array in the inner node is a pointer array that stores the child pointers. The array in the data node is a gapped array that stores key-payload pairs. Empty slots and key-payload pairs are interleaved, which reduces insertion cost overheads by requiring fewer shift operations to find the first available slot. A bitmap is used to identify an empty slot more easily.

To support lookup operations, ALEX traverses from the root to a data node using *model-based* search. When using model-based insertions, the predicted position of the inner node does not require any additional search process. When arriving at a leaf node, ALEX first calls the model to predict the location and then performs an exponential search to find the final position if needed.

To support insert operations, ALEX first uses a search process to find the slot where the key would be located. If the slot is already occupied, ALEX shifts items to obtain an empty slot for the new key. An SMO (Structural Modification Operation), which determines how to update the index structure, is triggered when a node is full. ALEX uses four mechanisms for an SMO, and provides a cost model for updating the tree structure.

**LIPP** [30]. LIPP has a single node type as shown in Figure 1(d). Each node in LIPP has a data array, a bit array, and a linear model. Each element in the data array can be one of the three types, *DATA*, *NULL*, and *NODE*. The bit array identifies the element type. A linear model predicts which slot to be accessed during a lookup. LIPP first adopts the Fastest Minimum Conflict Degree (FMCD) algorithm to obtain a linear model of a node with the smallest "conflict degree" (the maximum number of keys being inserted into the same slot). Then, LIPP inserts the key set using the resulting model in a single node. If only one key is inserted into a slot, this slot is labeled as *DATA* and stores a key-payload pair. If multiple keys are located in one slot, the slot is marked as *NODE* which stores a pointer to a child node. LIPP builds a new child node for any conflicting keys using the same process.

To support lookup operations, LIPP uses a linear model in each node to predict positions. If the slot is *NODE*, it accesses its children. If the slot is *DATA*, it checks if the key in this slot is the same as the lookup key – if true, it returns the payload; otherwise, it returns null. If the slot is *NULL*, it returns null.

To support insert operations, LIPP first performs a search to find which slot should hold the new key. If the slot is *NULL*, LIPP inserts the new key into that slot. If the slot is *DATA*, LIPP will create a new node for the inserted key and the key in that slot, mark the slot as a *NODE*, and store a pointer of the new node.

## 2.3 Other Updatable Learned Indexes

Model B+-tree [17] and RUSLI [20] are two recently proposed updatable learned indexes. A Model B+-Tree builds a model for each B+-tree node to predict which child node to access, and uses an update process similar to a B+-tree. If the predicted leaf node is not the target node based on a prediction error, a Model B+-tree will fetch more nodes than a B+-tree to locate the target leaf node, and which can be a significant overhead in certain cases. Therefore, we omit this index from our study. RUSLI extends RadixSpline [11] to support insertion by adding an overflow array. However, it has a very restrictive assumption – insertion keys must be drawn from a uniform distribution. Since such an assumption is rarely true in practice, we have not included this approach in our study.

Studies [4, 6] combine learned indexes with a log-structured merge (LSM) tree data structure [22]. A learned model is constructed for each SSTable (Sorted Strings Table), which is *immutable* after being created. Modifications (insert, update, delete) are supported in an LSM framework, and models are *rebuilt* during periodic compaction processes. Since they are implemented in a real production system, there are optimizations that we cannot reliably reproduce. PGM uses a similar idea to support insert (compaction), which we can test extensively. Thus, we exclude them from our study.

XIndex [27] and FINEdex [15] add concurrency support to the learned indexes. Wu et al. [31] propose a method based on the idea of Normalizing Flows [25] to transform the data distribution into an easier one to learn, and the proposed index structure is an extension of LIPP. CARMI [32] is a cache-aware learned index for main memory. Several recent studies have also adopted the idea of learned indexing on string data, spatial data, and multiple dimensional data [8, 16, 21, 24, 26, 28].

## 3 COMPARISON

As summarized in Table 1, different indexes introduce different design ideas based on one or more of the following aspects:

- *Searching an Inner Node.* ALEX and LIPP only use a model to predict the child and can be accessed in constant time. In contrast, B+-tree, PGM, and FITing-tree require a search of at least $O(\log m)$ time, where $m$ refers to the number of items in the node (B+-tree and FITing-tree) or an error bound (PGM).
- *Search on Leaf Node.* Except for LIPP, all of the indexing methods presented in Table 1 require a search stage (binary or exponential search) to find the exact position of a target key. The complexity is $O(\log m)$, where $m$ is the item count (in ALEX and B+-tree) or the error bound (in PGM and FITing-tree). Therefore, LIPP has the lowest cost to find a leaf node ($O(1)$ vs. $O(\log m)$).
- *Data Partitioning.* A data partitioning algorithm determines how many items are indexed in the inner nodes of the indexes. A smaller indexed item count produces lower tree heights and, in turn, faster search time. ALEX and LIPP partition the data into nodes using a learned model, namely *Model-based Insert*, which finds the exact position of a target key with no additional search.

- *Insertion.* (1) All indexes first use a search to find the position to insert a new key. Thus, insertions benefit from efficient search. (2) A FITing-tree (*Delta-Insert Strategy*) adds a "buffer" to hold new keys. B+-tree, ALEX, and LIPP nodes include extra space when creating nodes to store new items. In contrast, a B+-tree is a dense array, while ALEX and LIPP use gapped arrays. (3) If the target position of the new key is empty in a gapped array, ALEX and LIPP insert the new key in that position and finish the insertion process. However, a shift operation is always required in a B+-tree. If the predicted position is occupied by another key, ALEX shifts items to find an empty slot, and the gapped array can reduce the number of required shift operations. In contrast, LIPP creates a new node to hold the new key and any keys already occupying the predicted position to reduce future conflicts.

- *Structural Modification Operation (SMO).* When a "buffer" or node is full, all of the indexing methods must update the inner nodes in the tree structure. When updating leaf nodes, ALEX, LIPP, and a FITing-tree first fetch all the items and reinsert them into new nodes. When the node size is larger than that of a B+-tree, SMO can incur a much higher latency than a B+-tree.

## 4 LEARNED INDEXES ON DISK

We now show how to extend learned indexes to an on-disk scenario. Specifically, we use ALEX as a concrete example presented, followed by other indexes. We use ALEX as our example for the following reasons: (1) ALEX is the most difficult index to implement when all operations must be on-disk, due to the SMO requirement; (2) ALEX is a representative example which can be used to demonstrate common drawbacks in all existing updatable learned indexes when being ported to support on-disk operations.

### 4.1 Extending ALEX On Disk

We now discuss major extensions needed to implement ALEX to an on-disk configuration.

**Layout on Disk.** Figure 2 shows how to store the indexing data structure on-disk. All of the nodes in ALEX are stored contiguously. For each node, a model is stored, as well as utility structures (bitmaps, etc). In the original paper [7], the pointer array of one inner node stores the child pointers (each is an address to a memory position). When on-disk, we still need eight bytes to store child node addresses on-disk – 4 bytes for the block number and 4 bytes for the offset in the block. Since node sizes are variable, a node in ALEX may **cross multiple blocks**, especially data nodes (see *N5*). Multiple nodes can also be stored in one block (see *N6* and *N3*).

There are two different layout choices for ALEX: *Layout#1* in Figure 2(a) stores inner nodes and data nodes in the same file; *Layout#2* in Figure 2(b) creates one file for each type. Due to the small size of inner nodes in ALEX, one block in an *Inner Node File* can hold more than one node. Thus, we can traverse multiple levels using one block. For example, if we need to access *N5*, we need to fetch one block for each inner node (*N1* and *N3*) in *Layout#1*, but we only need to fetch one block in *Layout#2*. We implement both layouts for ALEX and test them using a lookup-only workload in Section 5. *Layout#2* has a 0.5%-30% performance improvement compared to *Layout#1*. Thus, we prefer *Layout#2* in our implementation.

Since the root node in ALEX can be changed as a consequence of an insert operation, the first block of the index is set as the *meta block*, which records the address of the root node. Additionally, a constraint is enforced such that the data in one node must be stored in an adjacent space. Otherwise, we need a mechanism to record the mapping between the blocks and nodes.

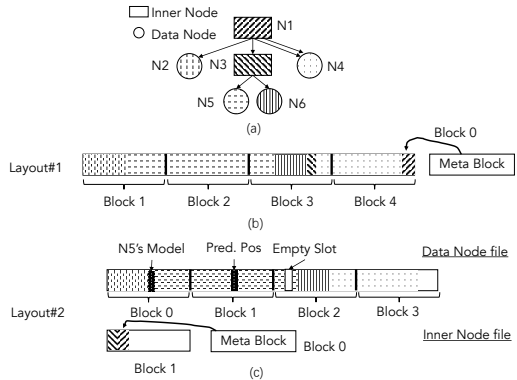

Fig. 2. The On-Disk ALEX Layout

**Query Processing on Disk.** When handling a search or an insert operation, first the block determined by the node model is fetched, and then the position in a node array is computed. Next, the block is accessed to obtain the on-disk child addresses. When the insert operation requires an SMO, new space is allocated for the new node and old nodes are marked as invalid.

When performing a scan operation (a range query), ALEX will first locate the smallest key in the search range and then scan forwards. A bitmap is used to skip empty slots. Since the size of the data node can be as large as 16 MB, the bitmap for one data node can cover at most 32 blocks if one block is 4 KB, and incurs additional I/O costs. Instead of loading all blocks for a bitmap into main memory, in our implementation, one block is loaded into main memory and scanned first. If the end of the scan range is found, we do not need to fetch any more blocks related to the bitmap.

ALEX also records several statistics, such as the number of shift and lookup-only queries. Hence, a write cost is incurred even for read-only queries. In our implementation, these records are not maintained for read-only queries.

Table 2. I/O Costs Analysis of the Studied Indexes on Disk.

|          | B+-Tree | ALEX | FITing-tree | LIPP | PGM |
|----------|---------|------|-------------|------|-----|
| Lookup | $\log_B N$ | $\log N + \log(M/B) + 1$ | $\log_B P + 2\epsilon/B$ | $2\log N$ | $\log(N/B)$ |
| Scan | $\log_B N + z/B$ | $\log N + \log(M/B) + z/B + 3$ | $\log_B P + 2\epsilon/B + z/B$ | $2\log N + z$ | $\log(N/B) + z/B$ |
| Insert | $2\log_B N$ | $(1 + 2M/B)\log N + 1 + \log(M/B)$ | $2\log_B P + 1 + 2M/B$ | $(2 + 2N/B)\log N$ | $\log(N/B)$ |

[1] $N$ is the total item count, $B$ is the maximum item count in one block, $M$ is the maximum item count in one data node (segment) of ALEX (FITing-tree), $P$ is the segment number in FITing-tree and PGM, $z$ is the item count in a scan (range query), $\epsilon$ is the predefined error bound in FITing-tree and PGM.

## 4.2 Extending Other Learned Indexes On Disk

For a FITing-tree, the *Delta Insert Strategy* is implemented and the following optimizations are included:

- A greedy segmentation algorithm is replaced by a more efficient streaming algorithm which was originally used in PGM [23].
- The original FITing-tree algorithm does not allow a key insertion if the key is smaller than the current smallest key. An extra buffer (one block) is introduced to hold keys such as this, and the physical address is recorded in the meta node. When this extra buffer is full, a segmentation algorithm is executed to partition the data and insert the new segments generated into the index.
- To support scan operations, additional metadata is added to the beginning of each segment to record the position of the left and right siblings, as well as how many items each of them contains. This is similar to the links between leaf nodes in a B+-tree.

We have also implemented an updatable, on-disk version of PGM. LIPP is similar to ALEX in that it is also an unbalanced tree structure with variable node sizes, and the on-disk layout is the same as ALEX – with the exception of the node bitmaps, which are replaced with a slot flag to identify the type. This removes the overhead of fetching the bitmap from the disk.

## 4.3 On-Disk I/O Cost Analysis

Table 2 shows the worst-case I/O cost of each index studied in an on-disk configuration.

**ALEX**. The tree height of ALEX is $\log N$ – the maximum fetched block count is $\log N$ when traversing to one data node. Because of the larger node sizes, there is a chance that the model and the slot required to access the node are not in the same block. So, there is *an extra block retrieval cost* when a data node is searched. The complexity of exponential search is $\log(M/B)$. Thus, the total cost of lookup is $\log N + \log(M/B) + 1$. ALEX uses the same process as a B+-tree to perform scan operations, but there are two extra blocks which store the bitmap for the node in ALEX. Thus, the complexity is $\log N + \log(M/B) + z/B + 3$. When performing an insert operation, ALEX first does a lookup to find a slot to insert a new key ($\log N + \log(M/B) + 1$). If the node is full, an SMO is required. ALEX reads all items in a node and constructs new nodes, which can propagate all the way to the root node ($(2M/B)\log N$). Thus, the total cost is $(1 + 2M/B)\log N + 1 + \log(M/B)$.

**FITing-tree**. A FITing-tree requires $\log_B P$ fetches from the root node to find the target segment, since the inner nodes of the FITing-tree are the same as a B+-tree. A binary search is invoked on the range of $2\epsilon$ items, which accesses $2\epsilon/B$ blocks in the worst case. The analysis of on-disk scans is the same as a B+-tree. To perform insertions, the *Delta Insert Strategy* is implemented, which introduces a sorted buffer to hold new keys. So, the cost is $\log_B P + 1$. If the buffer is full, the FITing-tree reads all items in the segment and buffer, and resegments them with a cost of $2M/B$. Once the new segments are added, the B+-tree component is updated. In the worst case, the cost is $\log_B P$ as all levels in the B+-tree may need to be updated.

**LIPP**. The height of a LIPP tree is $\log N$. Similar to ALEX, the larger node sizes of the upper nodes can result in a higher chance that the model and the slot are stored in different blocks. Thus, the lookup cost is $2\log N$. There is only one node type in LIPP, but different slot types. In the worst case, the item fetched is in multiple nodes and blocks. Thus, the total scan cost is $2\log N + z/2$. Similar to ALEX and FITing-tree, an insertion may lead to rebuilding a subtree with at most a height of $\log N$. At each level, at most $N$ items are read and written. Thus, the total cost is $(2 + 2N/B)\log N$.

**PGM**. The height of a PGM tree is $\log N$ and the error bound in PGM can be smaller than half of the maximum item count in a block. Thus, the complexity for search and scan is $\log N$ and $\log N + z/B$, respectively. For insert, we can only provide amortised time. In the worst case, PGM would merge all existing indexes into a new one as shown in Figure 1(b).

Table 3. Dataset Profiling under Error Bound and Conflict Degree (block size = 4 KB)

| Error | YCSB [7] | FB [18] | OSM [18] | Covid [29] | History [29] | Genome [29] | Libio [29] | Planet [29] | Stack [29] | Wise [29] | OSM(800M) [18] |
|---|---|---|---|---|---|---|---|---|---|---|---|
| 16 | 70,135 | 2,120,485 | 1,351,170 | 231,852 | 303,737 | 3,153,966 | 291,257 | 1,416,012 | 54,073 | 246,463 | 6,175,387 |
| 64 | 6,952 | 523,006 | 326,932 | 42,695 | 40,817 | 295,604 | 77,401 | 268,247 | 6,956 | 27,553 | 1,375,143 |
| 256 | 23 | 119,891 | 81,392 | 8,630 | 8,464 | 23,228 | 19,333 | 55,061 | 950 | 4,713 | 328,623 |
| 1024 | 1 | 18,495 | 20,925 | 1,890 | 2,029 | 4,975 | 3,616 | 12,001 | 196 | 1,184 | 81,577 |
| B+-tree | 980,393 | 980,393 | 980,393 | 980,393 | 980,393 | 980,393 | 980,393 | 980,393 | 980,393 | 980,393 | 3,921,569 |
| Conflict Degree | 4 | 114 | 4,106 | 27 | 9 | 585 | 2 | 22 | 1 | 10 | 10,107 |

## 5 EXPERIMENTAL SETUP

### 5.1 Datasets & Profiling

**Datasets.** We use eleven datasets, which are widely used in existing studies [7, 18, 29, 30]. The first ten have 200M keys and each key is an *uint_64* integer. We use the payload as the key plus 1. The dataset size is 2.98 GB in each of the first ten datasets. The last one has 800M keys with 11.92 GB size used for scalability experiments.

**Profiling.** As shown in Section 2, all learned indexes use a linear function as the model. If a dataset is difficult to model with a linear function, the performance of learned indexes can be degraded.

- For FITing-tree, PGM, and ALEX, we use the segmentation algorithm from PGM and several *error bound* settings to show how hard it can be to model certain datasets using a linear function. *A dataset with more segments is harder to model under the same error bound.* The results are shown in Table 3.
- The performance of LIPP is related to the *conflict degree* of a dataset [30], which is shown in the last row of Table 3. *A dataset with a larger conflict degree lowers performance for LIPP.*

We also report the leaf node number in a B+-tree when the block size is 4 KB. For ALEX, PGM, and a Fitting-tree, we set the default error bound to 64 – with the worst performing dataset being *FB* for this error bound. Under this metric, *OSM* is the most difficult one. While on both settings, *YCSB* is the easiest one.

Due to space limit, we report the performance using three representative datasets – *FB*, *YCSB*, and *OSM*. For the remaining datasets, readers can refer to a technical report [3].

### 5.2 Workloads

We test all the indexes on six different workload types: (1) **Lookup-Only** workload, which performs lookups on the indexes built by bulkloading all keys in each dataset. We randomly sample 200,000 lookup keys from the existing keys. (2) **Scan-Only** workload, which performs scan operations on the same indexes as the Lookup-Only workload. A scan operation is implemented with a lookup operation on the start key and a scan of the next 99 elements. The start keys are generated in the same way as the Lookup-Only workload does. (3) **Write-Only** workload, which inserts 10M key-payload pairs in the indexes after bulkloading 10M random keys. (4) **Read-Heavy** workload, performs 90% lookups and 10% inserts on indexes after bulkloading 10M random keys, i.e., we perform 2 inserts and 18 lookups, then repeat the process. (5) **Write-Heavy** workload, which performs 90% inserts and 10% lookups on indexes after bulkloading 10M random keys. We perform 18 inserts and 2 lookups, then repeat the process. (6) **Balanced** workload, which performs 50% inserts and 50% lookups on the indexes after bulkloading 10M random keys. We perform 10 inserts and 10 lookups, then repeat the process.

In each of the Mixed workloads (Read-Heavy, Write-Heavy, Balanced), the total number of operations is 10M and the search keys for the lookup in the Mixed workloads are evenly distributed.

### 5.3 Other Implementation Details

**Code & Environment.** We implement all the indexes in C++ and make them available at [1]. We conduct the experiments on a HDD using Red Hat Enterprise Server 7.9 on an Intel Xeon CPU E5-2690 v3 @ 2.60GHz with 256 GB memory and a 1TB HDD, and the experiments on SSD using Ubuntu 20.04 on AMD EPYC 7662 with 500 GB memory and four 8TB SSD.

**Metrics.** (1) The storage size of the whole index used on disk; (2) average throughput for each workload type and tail latency for the lookup-only workload and write-only workload; (3) average block count per search query (lookup-only and scan-only workloads).

**Parameters.** For a FITing-tree, we set the buffer size of each segment to 256 and the error bound to 64 by default. Since the optimal error bound setting for a FITing-tree can vary across datasets and workloads, we test several different error bounds and find that the FITing-tree achieves good performance in the majority of test cases when the error bound is set to 64. PGM, ALEX and LIPP use the default parameter settings from the original papers. In Section 6.4, we perform an experiment that explores the impact of block size. For other experiments, we fix this value to 4 KB.

## 6 EXPERIMENTAL EVALUATION

Our comprehensive evaluations in Section 6.1-6.6 aim to answer:

**Q1:** How good are learned indexes when compared to a B+-tree on an HDD and an SSD, if the entire index structure is disk-resident?

**Q2:** Is there any benefit to storing inner nodes of the learned indexes in main memory?

**Q3:** How much storage do learned indexes require?

**Q4:** What impact do different block sizes have on performance?

**Q5:** Do learned indexes have robust performance when disk-resident?

**Q6:** What impact does the buffer have?

### 6.1 Evaluation When the Entire Index is Disk-Resident

**Setting.** In this set of experiments, we assume that the meta block, which records the root node address and other utility information, is stored in main memory when in use, while the remaining index structure remains disk-resident, and stored within blocks of size 4 KB. There is no buffer management, i.e., for each request, we must read/write the required blocks from disk.

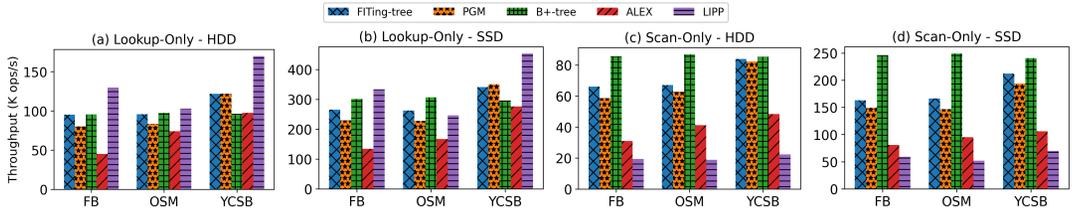

Fig. 3. Search Performance Comparison on HDD and SSD: the entire index is disk-resident using blocks of size 4KB.

Table 4. An Analysis of Fetched Block Counts for Lookup-Only and Scan-Only Workloads. For LIPP, its inner node count for the Scan-Only workload is provided in the brackets.

|  | FB | | | | OSM | | | | YCSB | | | |
|---|---|---|---|---|---|---|---|---|---|---|---|---|
|  | FITing-tree | PGM | ALEX | LIPP | FITing-tree | PGM | ALEX | LIPP | FITing-tree | PGM | ALEX | LIPP |
| Inner Node Count | 3 | 5 | 6.7 | 1.8 (18.8) | 3 | 5 | 2.7 | 2.3 (23.1) | 2 | 3 | 3 | 1.3 (16.7) |
| Inner Block Count | 3 | 3.9 | 6.5 | - | 3 | 3.7 | 2.6 | - | 2 | 2 | 2.2 | - |
| Leaf Block Count (Lookup) | 1.2 | 1.3 | 2.6 | 3.0 | 1.2 | 1.2 | 2.2 | 3.8 | 1.2 | 1.3 | 2 | 2.3 |
| Leaf Block Count (Scan) | 2 | 1.7 | 4.1 | 24.0 | 1.8 | 1.5 | 3.8 | 30.0 | 1.6 | 1.7 | 3.6 | 19.7 |

*6.1.1 Lookup-Only Workload.* Figure 3(a) and Figure 3(b) present the throughput of an HDD and an SSD, respectively. Figure 4(a) reports the average number of fetched blocks per query. We first present the observations (**O**) and then provide a detailed analysis.

**O1: When the entire index is disk-resident, the throughput of the Lookup-Only workload is determined by the number of fetched blocks from disk.** The increase in the number of fetched blocks usually degrades performance, since fetching data from disk tends to dominate the execution time. For example, the B+-tree and FITing-tree show similar performance on the *FB* and

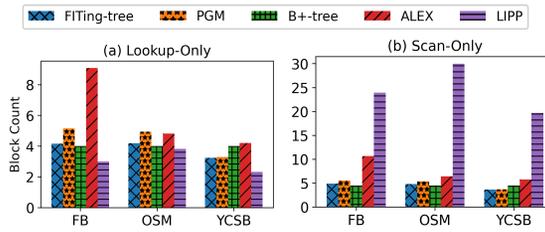

Fig. 4. Average Fetched Block Count for Search on HDD.

*OSM* datasets while the FITing-tree has a much higher throughput on *YCSB* (26.2% improvement for an HDD).

**O2: Most existing learned indexes are competitive or outperforming B+-tree on the Lookup-Only workload.** LIPP outperforms all other indexes on our test datasets. When compared to the alternatives, the node fanout of LIPP is much larger, which contributes to its lower tree height and leads to fewer fetched blocks. However, a large fanout leads to a larger index size, which implies longer index construction time, as shown in Figure 7(a) and 7(b). ALEX performs the worst on *FB*, which is attributed to the highest number of fetched blocks.

**O3: On the Lookup-Only workload, B+-tree exhibits stable performance, while the performance of the learned indexes fluctuates.** This is because learned index throughput depends on the difficulty of modeling the data distribution. For LIPP, *OSM* is the most difficult dataset, while *FB* makes the most difficult dataset for ALEX, FITing-tree, and PGM. Since the B+-Tree does not require any predictions from a linear model, its performance remains consistent across all the datasets.

**Analysis of Fetched Blocks for the Lookup-Only Workload.** We breakdown the fetched block counts for each index into two components: (a) the inner nodes, and (b) the leaf nodes. The results are presented in Table 4. Since there is only one node type in LIPP, we report the average total node count for LIPP. On the three datasets tested, the B+-tree has 4 levels in all cases, with 3 inner node blocks and 1 block for the leaf node.

Based on Table 4, we observe that: (1) For the FITing-tree, accessing one inner node will fetch one block; in contrast, in the case of PGM and ALEX, more than one inner node can be stored within a block due to the small node sizes at the upper tree levels (hence inner block count is sometimes smaller than the node count for those). (2) Compared to the B+-tree, the FITing-tree and PGM have a smaller search range, 256 vs. 128. However, the average fetched block counts in the FITing-tree and PGM are slightly larger than the B+-tree (by 1 block). This is because the FITing-tree and PGM cannot guarantee the entire search range to be stored in one block. (3) Compared to the other indexes, ALEX accesses more leaf node blocks. Due to the large node size of the leaf nodes, there is a greater chance that the model stored in the node header resides in a different block than the predicted target position. Thus, ALEX reads at least 2 blocks. Moreover, on *FB* and *OSM*, the fetched block count is larger than 2. Similar to the binary search in the FITing-tree and PGM, exponential search in ALEX can occur across multiple blocks. (4) Although LIPP accesses fewer nodes than the other indexes, on average, it requires more than 1.65 fetched blocks per level. The large node size in LIPP can cause the model and the predicted position to reside in different blocks.

*6.1.2 Scan-Only Workload.* From Figure 3(c)-(d), we observe that:

**O4: For the Scan-Only workload, regardless of the dataset hardness, B+-tree outperforms others across all datasets.** Just as observed in **O1**, the throughput of the learned indexes is highly dependent on the number of fetched blocks.

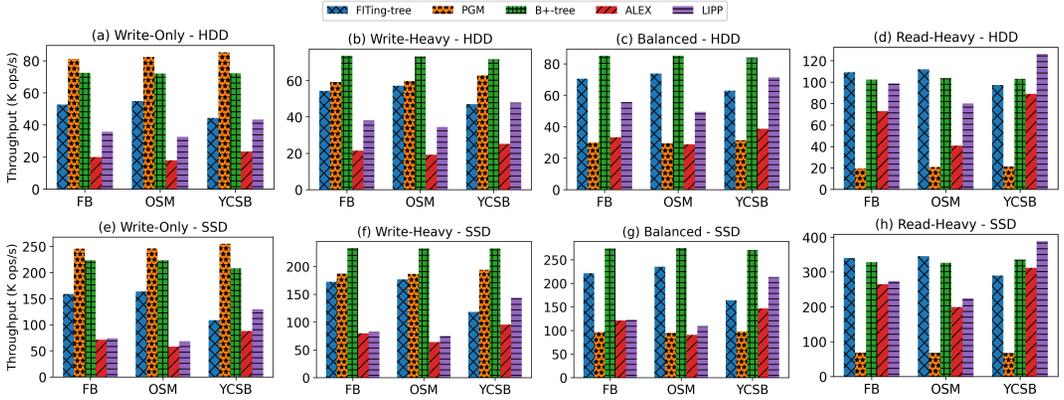

Fig. 5. Write Performance Comparison on HDD and SSD: the entire index is disk-resident using blocks of size 4KB.

Table 5. An Average Fetched Block Count for Search on HDD under the Hybrid Design. The first (second) number is the block count for the Lookup-Only (Scan-Only) workload.

|      | FITing-Tree | PGM       | ALEX      | LIPP     | B+-tree  |
|------|-------------|-----------|-----------|----------|----------|
| FB   | 3.25/3.74   | 3.25/3.74 | 4.02/4.51 | 3.15/3.64| 4.0/4.49 |
| OSM  | 4.25/4.74   | 4.17/4.66 | 4.77/5.26 | 4.5/5.0  | 4.0/4.49 |
| YCSB | 3.25/3.74   | 3.25/3.74 | 4.0/4.49  | 3.01/3.5 | 4.0/4.49 |

**O5: ALEX and LIPP exhibit the worst performance in the Scan-Only workload.** This is attributed to the fact that ALEX and LIPP fetch many more blocks compared to the others.

**Analysis of Fetched Blocks for the Scan-Only Workload.** To support scan queries, all the indexes first locate the position of the start key and then scan forward until they reach the final key. We set the start keys for the Scan-Only workload with the keys from the Lookup-Only workload. Thus, the inner node counts and inner block counts are identical to the Lookup-Only workload for FITing tree, PGM, and ALEX as presented in Table 4. In the last row, we report the fetched block counts at the leaf node for scan queries. For LIPP, we also report the total fetched node counts and block counts as presented in the brackets.

From Table 4, we observe that: (1) LIPP fetches the highest number of nodes. The three entry types (i.e., *NODE*, *DATA*, *NULL*) are interleaved in one LIPP node. If we locate a *NODE* entry, we must fetch a new node from disk and scan it. If we visit all elements for a node and its children, we traverse back to the parent node and continue the scan. There is a high chance that nodes are stored across multiple blocks, which leads to the highest fetched block count. Furthermore, the elements in a node can be stored across multiple blocks, which also results in an increased fetched block count. (2) To skip the empty slots in the data nodes, ALEX introduces a bit array to indicate if a slot is empty. The bit array may require additional blocks to be fetched. In contrast, B+-tree, PGM, and FITing-tree, store the key-payload pairs contiguously, and the sibling segments (leaf nodes) are linked. Therefore, there is negligible or no overhead in fetching the utility data or traversing nodes. **How Good a Hybrid Index Structure Design Is.** Storing key-payload pairs continuously benefits scan in B+-tree, FITing-tree, and PGM. Thus, an emerging idea is to maintain the leaf nodes in a continuous manner as in a B+-tree, and adopt learned indexes as the inner part to index the maximum keys in leaf nodes. Following this idea, we present the average fetched block count for the Lookup-Only and Scan-Only workloads in Table 5 and make several observations: (1) On *FB* and *YCSB*, all hybrid designs achieve similar or better performance than a B+-tree. When we only index the maximum keys in the leaf nodes, *OSM* is a harder dataset than *FB* and *YCSB*. The segment counts under an error bound of 64 are 107, 1, 1457 in *FB*, *YCSB*, and *OSM*, respectively. (2) With a B+-tree-styled leaf node, ALEX and LIPP perform better on scan compared to the original design. (3) LIPP performs the best on *YSCB* and *FB*. However, compared to the original LIPP (in Table 4), the fetched block count in the Lookup-Only workload is a little larger. With the maximum keys in each leaf node in the LIPP part. We have to scan forward to find the next *DATA* slot if meeting a *NULL* slot.

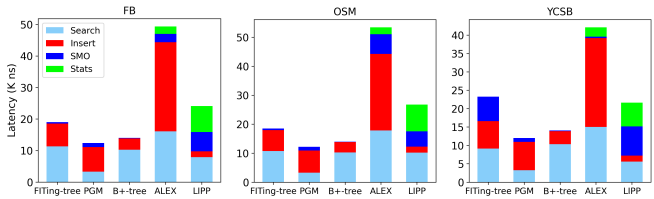

Fig. 6. Write Performance Breakdown.

### 6.1.3 Write-Only Workload.
The results presented in Figure 5(a) and Figure 5(e) lead to the following observations:

**O6: For the Write-Only workload, the relative ranking of all indexes is consistent across all datasets, with PGM significantly outperforming other methods.** PGM supports arbitrary insertions using an LSM style tree, which uses a small sorted array (of fixed size) to cache new insertion requests. When the sorted array is full, PGM merges this data into the static PGM index (increasing its size), as shown in Figure 1(b). Therefore, most insertion requests need to read and write only a smaller number of blocks.

**O7: Other than PGM, B+-tree significantly outperforms other learned indexes on the Write-Only workload.** All of the indexes first find the slot to store a new key-payload pair. Although learned indexes have shown similar or better performance than B+-tree on the Lookup-Only workload, the overhead for insertion outweighs the benefit acquired in the search process. The performance of ALEX and LIPP is severely impacted in this case. In order to support SMO, LIPP maintains statistics for each node. Thus, for each insert, LIPP will update all of the nodes in the path to the inserted node. ALEX needs to maintain such statistics for leaf nodes as well.

**O8: B+-tree and PGM exhibit consistently good performance across all tested datasets, while the performance of other learned indexes fluctuates significantly.** Overall, LIPP performs the worst on the harder datasets. This is similar to ALEX that exhibits lower throughput in harder datasets. Interestingly, FITing-tree performs the worst in the easier datasets.

**<u>Performance Analysis for the Write-Only Workload.</u>** We break the insert process into four steps: (a) initial search step, to find the insert position, (b) insertion step, to do the insertion, (c) SMO step, to do structural modification, and (d) maintenance step, to update statistics related to the SMO. We report the average latency of each step in Figure 6. Unlike other indexes, PGM initiates the search process to find the location of a new key-payload pair in a sorted array of size 585 (3 blocks). PGM only needs to fetch one or two blocks to find the position. FITing-tree has similar search time to B+-tree, but exhibits larger insertion time, since FITing-tree needs to write an extra block to update the current item count for a segment.

ALEX has the highest latency for the insertion step due to several reasons: (1) It reads from the extra blocks for the bitmap to check if the predicted position is empty, and then updates the bitmap after the new insertion; (2) If a predicted position is not empty, a shift operation is triggered to create an empty slot for the key-payload pair, which may move the items across blocks; (3) Although the bitmap is used to determine if a slot is empty, ALEX will still overwrite empty slots until reaching the previous element to avoid accessing the bitmap during the lookup query. In (2) and (3), ALEX also needs to read the bitmap to find any related empty slot(s).

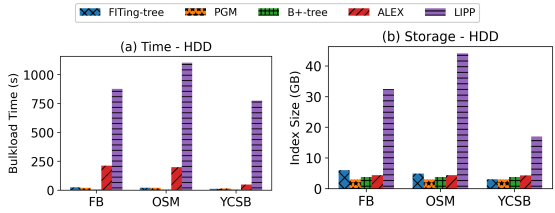

Fig. 7. Bulkload Performance on HDD.

The FITing-tree in *YCSB* has a larger SMO overhead compared to *FB* and *OSM*. *YCSB* can be modeled using fewer models and has more items in each segment. When triggering an SMO operation, the FITing-tree will write more blocks for *YCSB*. ALEX and LIPP have a moderate SMO overhead for *FB* and *OSM*. For ALEX, a harder dataset induces more SMO operations. For example, ALEX requires 57, 330,236, and 212,917 SMO operations for *YCSB*, *FB*, and *OSM*, respectively. LIPP has two types of SMO, creating a new node to eliminate the conflict, and adjusting the tree structure. In our testing, millions of SMOs are the second cases, i.e., an SMO for every three insertions. Interestingly, LIPP has a larger SMO overhead on *YCSB*, an easy dataset. This is because many keys are conflicted locally.

LIPP also has a larger maintenance overhead compared to ALEX. This is because LIPP will update all nodes in the access path, while ALEX only needs to update one data node for the insertion.

### 6.1.4 Read-Write Workload.
From Figure 5(b) - (d) and Figure 5(f) - (h), we derive the following observations:

**O9: In most cases (6 out of 9), the B+-tree outperforms learned indexes. Since learned indexes incur a higher number of block writes, when disk-resident, learned indexes have subpar performance compared to the B+-tree (which ranks first or second in all the experiments with this workload). The second case rarely happened in our testing while there are** $4.28M$**,** $3.36M$**, and** $2.81$ Excluding PGM, the benefit of the lookup gained by the learned indexes is eclipsed by the overhead created by the insertion operations. For example, LIPP outperforms others in Lookup-Only workloads, yet it exhibits worse performance than the B+-tree for *FB* and *OSM*, even in the Read-Heavy workloads.

**O10: As the read ratio increases, the throughput of PGM degrades severely, unlike the alternatives where it increases.** As discussed in **O6**, PGM performs well for insertion operations. However, due to the LSM tree layout, PGM must maintain several static PGM indexes (see Figure 1(b)). To support a read query, PGM must access all static PGM indexes sequentially until finding the search key, or determining that the key does not exist. Each static index is stored as a separate file – leading to a higher I/O cost.

### 6.1.5 Bulkload.
From Figure 7(a) and Figure 7(b), we observe that:

**O11: Learned indexes usually require more storage space than B+-tree, and all of them require more time to build an index on disk.** On all datasets, PGM has the smallest index size, while LIPP has the largest index size. PGM supports insertions using LSM. Each static PGM index does not need to allocate extra space to hold the new insertions. On the contrary, the other learned indexes need to allocate extra space before insertions occur. FITing-tree allocates a fixed size buffer for each segment. ALEX and LIPP use a gapped array. In LIPP, if the count of items (*item_count*) inserted into a single node is in the range [100,000, 1,000,000), LIPP allocates 2∗*item_count* slots. If the count < 100,000, 5∗*item_count* slots are allocated. LIPP has the largest empty slot ratio compared to others.

**O12: The index sizes of FITing-tree and LIPP highly depend on the distribution of the dataset indexed.** For harder datasets, more segments are generated by the FITing-tree, and more buffers are preallocated. A potential way to optimize the storage size of the FITing-tree is to allow lazy buffer allocation, where a buffer is only allocated for a segment when a new key-payload pair must be inserted into it. If more than one item is inserted into a slot in LIPP, a new node is created. Harder datasets usually create more nodes. A high empty slot ratio (as discussed in **O11**) in new nodes increases the storage size.

## 6.2 Evaluation When the Inner Nodes are Memory-Resident

**Setting.** In this section, we investigate the impact of indexes when inner nodes are memory-resident. For the B+-tree and FITing-tree, we use an STX B+-tree [2] for the inner nodes, which is also adopted by ALEX and LIPP as their B+-tree baseline. For PGM and ALEX, we use the original implementation for the inner nodes and store the leaf nodes on disk. We exclude LIPP from this experiment for two reasons: (1) There is only one node type in LIPP; and (2) The largest node in LIPP is the root node. Even if only the root node is stored in main memory, it requires more than 3 GB of space for the *FB* dataset, while other indexes require no more than 40 MB space. We refer to this setting as the *hybrid case* in the rest of the paper.

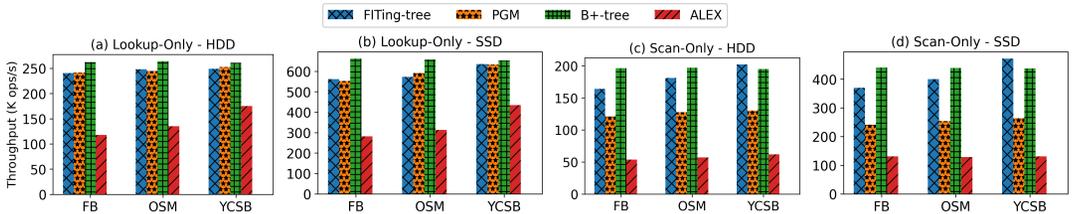

Fig. 8. Search Performance Comparison on HDD and SSD: the inner nodes are memory-resident, leaves are on disk.

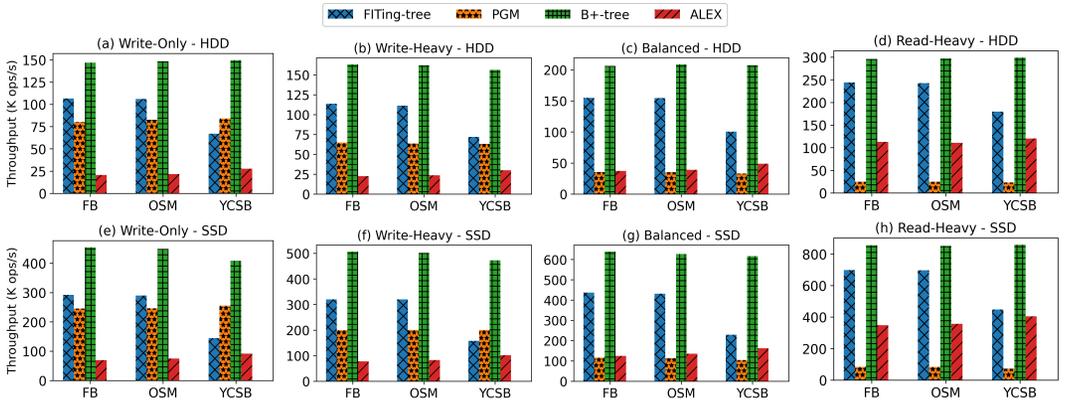

Fig. 9. Write Performance Comparison on HDD and SSD: the inner nodes are memory-resident, leaves are on disk.

### 6.2.1 Lookup-Only Workload & Scan-Only Workload. When inner nodes are memory-resident, the number of fetched blocks for each index is the same, as shown in the last two rows in Table 4. Given Figure 8 and Table 4, we observe that:

**O13: FITing-tree and PGM are competitive with B+-tree. However, ALEX is not.** Just as what we saw in the on-disk case, performance is determined by the number of blocks fetched.

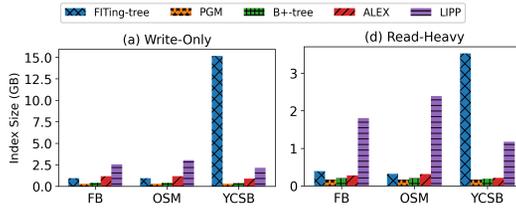

Fig. 10.   Storage Usage on Disk: the entire index is disk-resident using blocks of size 4KB.

Any performance gains from faster lookups in main memory are eclipsed by on-disk reads, which remains the key bottleneck.

*6.2.2   Workloads With Write Queries.* For workloads requiring write operations presented in Figure 9, we observe that:

**O14: Unlike the other indexes, utilizing main memory to store inner nodes does not help PGM.** Other indexes first search for a slot to hold a new key-payload pair. Thus, the performance gains of lookup-only queries in the hybrid case can improve write performance. However, for PGM, the write process is the same as that on disk, so only marginal performance gains can be observed. One idea to alleviate the problem is to keep the sorted array of PGM in main memory, as is done for the skip lists of LSM based systems.

**O15: B+-tree outperforms the others across all workloads in the Hybrid case.** From Figure 6 we observe that the main overhead for ALEX and LIPP when they are disk-resident is for insertions, structural modifications, and statistic updates. Therefore, putting the inner nodes in main memory does not help, since they still need to pay this cost. B+-trees have smaller insertion cost than PGM. After bringing inner nodes into main memory, B+-tree has better performance in the initial search step, which helps it outperform PGM. Similarly, the decrease of search overhead in the FITing-tree means it has better performance than PGM for *FB* and *OSM*. However, due to large SMO overheads for *YCSB*, the FITing-tree is worse than PGM for *YCSB*.

### 6.3   Storage Size Study

Unlike the main memory setting, the memory allocated on disk cannot be reclaimed directly and we do not consider how to reuse the invalid space here[1]. The result is reported in Figure 10. Other workload types show a similar pattern to the Write-Only workload.

**O16: Other than FITing-tree for *YCSB*, the storage size ranking of the studied indexes is the same as presented in the bulkload experiment.** Although we use the same dataset sizes for *FB* and *YCSB*, FITing-tree has a larger storage size for *YCSB*, since *YCSB* can be modeled using fewer linear models. That is, each segment covers more keys than in *FB* (more than 5,000 key-payload pairs, or a total of 19 blocks). SMO operations allocate more blocks in the FITing-tree than in any other index. Interestingly, the FITing-tree has to write more blocks for *YCSB*, but still requires fewer than ALEX, and is competitive with LIPP for *YCSB* (see Figure 5). Given the performance breakdown shown in Figure 6, the overhead required to update statistics and do insertion eclipses the performance gains achieved by fewer SMO operations.

PGM and B+-tree have much smaller storage sizes. When splitting a node of a B+-tree, a small part of the original node is retained, and the rest goes to a new node. In contrast, learned indexes use learned models to predict positions, therefore we *must store this as part of a node contiguously on-disk*. This makes it difficult to reuse reclaimed space, and hence higher fragmentation can be observed. For PGM, when a smaller sized index has been merged, the corresponding index file can be deleted from disk.

---

[1]One way to alleviate the issue is to reuse the invalid space by bookkeeping the address and size of invalid space. However, this process incurs extra overhead to maintain the information for invalid space

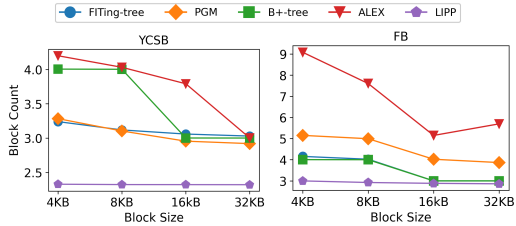

Fig. 11. Fetched Block Count under Different Block Sizes.

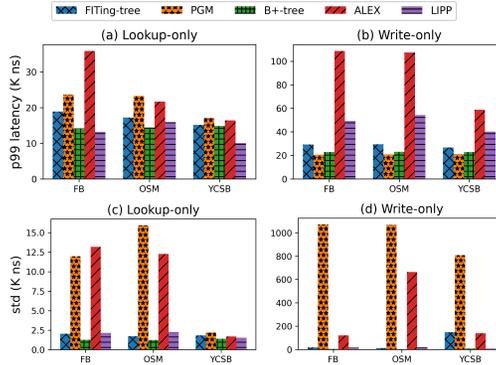

Fig. 12. Tail Latency of the Lookup-Only Workload and Write-Only Workload on HDD.

## 6.4 Impact of Block Size

In Figure 11, we show the performance of the Lookup-Only workload when the block size is varied on a HDD. We observe that:

**O17: The block size variation has different impacts on different indexes.** The number of blocks fetched by LIPP does not change when the block size is varied. The position prediction in LIPP is accurate and requires no additional search operations. The physical address of a target element can be derived from the start address in a node, and the relative position from the model prediction. For the other indexes, larger block sizes usually lead to fewer blocks fetched. When the block size is 16 KB, the height of B+-tree decreases for a larger fanout. This is also why FITing-tree's performance changes on *FB*. For FITing-tree and PGM, the error bound = 64, and the length of search range = 128. There is a higher chance that elements in the search range will be in the same block when the block size is large, which also reduces the blocks fetched in leaf nodes for ALEX using an exponential search. In ALEX and PGM, larger block sizes increase the storage available to nodes in the upper levels, so we can traverse more nodes using one block.

## 6.5 Tail Latency

In Figure 12, the tail latency (99-th percentile latency and standard deviation) for the Lookup-Only workload and Write-Only workload are reported. We make the following observation:

**O18: Learned indexes usually have greater p99 latency than the B+-tree for Lookup-Only and Write-Only workloads, and exhibit less stable performance.** The 99-th latency performance ranking is the same as observed in Figure 3(a) and Figure 5(a). Interestingly, for the Lookup-Only Workload, PGM and ALEX have a larger standard deviation on *FB* and *OSM*. PGM uses a balanced tree structure, so it should have a similar query latency. We record the number of fetched blocks for each test query. Given the constraint that a binary search is conducted only on

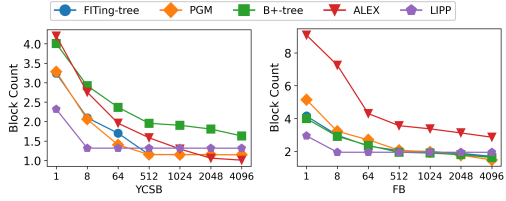

Fig. 13. An Average Fetched Block Count from Disk under Different Buffer Sizes on the Lookup-Only Workload. Buffer size indicates how many blocks can be cached.

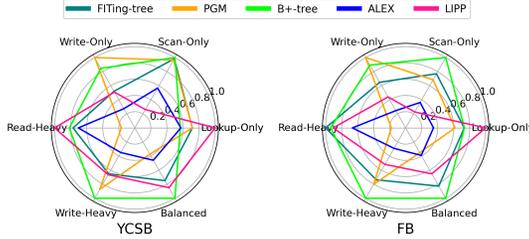

Fig. 14. Comparison of ALL Workloads on *YCSB* and *FB*. The entire index is disk-resident and each index's throughput is normalized by the largest one in the corresponding workload (the higher is the better).

one block at a time, in the worst case, PGM may have to access two blocks alternately. As discussed in Section 6, we do not include buffer management by default; instead, we check whether the last block fetched can be reused. ALEX is on the other hand an unbalanced tree structure. Thus, different data regions may have different query latencies. If the insertion requires an SMO, it will have a larger latency, leading to a larger standard deviation. That is why ALEX exhibits a larger standard deviation for *OSM*, and FITing-tree has a greater variance for *YCSB*. LIPP frequently issues an SMO to create a new node (in every three insertions), and shows a low variance for the writes.

## 6.6 Buffer Size Study

We vary the number of blocks that can be cached in the main memory and test on the Lookup-Only workload. We adopt the LRU strategy. From Figure 13, we make several observations: (1) With zero or only a small buffer size, LIPP has the smallest fetched block count due to its lowest average tree height. (2) With a larger buffer size, the other indexes will fetch fewer blocks from the disk. When the buffer size is larger than 8, LIPP is outperformed by the other indexes. One reason is that LIPP has a larger node size in the upper level, leading to a high probability of fetching blocks that are not in the buffer. (3) PGM achieves the best performance with a large buffer size due to its small non-leaf nodes' size.

## 7 INSIGHTS AND DESIGN CHOICES ON DISK

### 7.1 Shortcomings of Learned Indexes on Disk

Figure 14 presents the overall performance of all the studied indexes on *YCSB* and *FB*. *Except for Lookup-Only Workloads, the B+-tree is either competitive or outperforms learned indexes on disk.* We now summarize commonly observed shortcomings (**S1 - S5**).

**S1 – Model-Slot Overhead.** Before accessing a slot in a node, ALEX and LIPP need to fetch the learned model in a node. The model is stored in the node header. The model and the searched slot may be located in different blocks. Although learned indexes can have lower tree heights, the extra blocks needed to fetch the model can lead to more blocks being accessed overall compared to the B+-tree. This problem does not exist neither in the FITing-tree nor PGM, since they store the model in the parent node.

**S2 – Search Overhead.** Even for very small search ranges where FITing-tree, PGM and ALEX routinely outperform B+-tree in main memory, it is not the same case on disk, because no learning approach can guarantee that (even very few) searched items will be stored in the same block. Without judicious buffer management developed for learned indexes, this problem may be severe in extreme cases. Moreover, PGM supports arbitrary insertions but the LSM structure must maintain multiple files. When handling a query in mixed workloads, PGM may access multiple files, and hence exhibits much worse performance than others. LIPP does not have this problem since it tends to have accurate predictions in each step. However, since LIPP does not distinguish between the data and inner nodes, traversal of additional nodes may still be required.

**S3 – Utility Structure Overhead.** This overhead comes from two parts. (1) For searches – when performing a scan operation (a range query), ALEX will first locate the smallest key of the search range and then scan forwards. The bitmap used by ALEX can skip empty slots. However, due to the size of the data nodes (up to 16MB), the bitmap for one data node can cover at most 32 blocks, and incur additional I/O costs. (2) For writes – both ALEX and LIPP maintain statistics based on historical workloads which are leveraged in SMO operations. These statistics are stored in the node header. Thus, ALEX and LIPP must update the header of the node after each insertion. From Figure 6, we can see that the overhead for this operation is far from negligible when the index is disk-resident.

**S4 – SMO Overhead.** As shown in Figure 6, ALEX, LIPP, and FITing-tree have large SMO overheads in certain cases. The SMO operation is required in order to boost performance of the later queries. For example, this operation reduces the tree height in LIPP. On the other hand, this maintenance overhead can hurt the overall performance, especially for on-disk operations (see Figure 5 and Figure 9). The B+-tree thus outperforms all of them on nearly every workload involving a write operation.

**S5 – Other Index-Specific Overheads.** (1) Although the gapped array in ALEX helps reduce shift operations, supporting lookup queries without accessing a bitmap forces ALEX to overwrite the preceding empty slots until it reaches the previous element. With a large leaf node size, ALEX may have to update more blocks than the B+-tree. (2) If the predicted slot is occupied by another key-payload pair, ALEX will shift items to obtain an empty slot. This may move items across blocks, leading to extra block writes.

## 7.2 Design Choices

Based on the above observations, we propose the following best practices when designing disk-resident learned indexes:

**P1 – Reduce the tree height.** The tree height directly influences the number of fetched blocks. Designing smaller-sized nodes may be an alternative choice. This implies that multiple nodes can be stored in the same block. The latter will benefit search-only workloads. However, after updating the nodes, some of them may need to be moved to another block. LIPP and the interpolation search tree [5, 19] can reduce the tree height while they only benefit lookup-only queries and can increase the storage cost. Moreover, as shown in Section 6.6, LIPP cannot utilize the buffer well due to its large node size at upper level. A potential solution worth investigating is to combine it with another index structure.

**P2 – Use a light-weight structural modification operation (SMO).** Although existing learned indexes have the advantage of faster search, higher overheads stemming from SMO operations reduce the performance for the Write-Only and Mixed workloads. There are three aspects we should consider: (1) reducing the frequency of index tree modifications; (2) reducing the size of the partial trees that must be updated during each structure modification operation; (3) reducing the reliance on historical statistics in the SMO process or storing this information so that the process requires far fewer block fetches or updates.

**P3 – Lower overhead when fetching the next item.** Although the design of gapped array boosts the performance of ALEX and LIPP for insert, it has an additional overhead–more blocks are fetched when skipping empty slots on disk. Without the ability to differentiate between data nodes and inner nodes, LIPP cannot quickly locate the next item needed, and traverses additional nodes. When using learned indexes on disk, key-payload pairs should be stored contiguously, or alternatively, the nodes that store the key-payload pairs should have higher densities. As shown in Section 6.1.2, using a B+-tree styled leaf node may mitigate this issue.

**P4 – Storage layout design.** Larger node sizes in learned indexes reduce the height of the tree structure in the main memory configuration. However, when we store the larger nodes on disk, there is a much greater chance that two blocks must be fetched to locate the searched slot – one for fetching the model and one for fetching the address of the child node (or the data). If we cannot reduce the tree height by 2 levels for each large node that crosses multiple blocks, the learned indexing on-disk becomes less attractive. A possible optimization is to store the model in the parent node (similar to the idea adopted by PGM and FITing-tree), which will result in fetching at most one block for each level traversed.

**P5 – Co-design learned index when using buffer.** Co-optimizing the size of the in-memory structure (inner nodes) combined with an additional search cost over on-disk leaf nodes is a plausible strategy. A B+-tree styled leaf node may be used, which usually results in fewer fetched blocks compared to the original learned index, and is also efficient for scan-based queries, as shown in Section 6.1.1. When designing a learned index on disk in this manner, the performance difference depends on which learned index is used in main memory. Based on a recent study [29], ALEX, LIPP, and ART [14] would be suitable choices. Another plausible strategy is to cache frequently accessed blocks. As shown in Figure 13, indexes with a smaller size at the upper level, e.g., PGM, ALEX, B+-tree, and FITing-tree, are more suitable for larger buffer sizes.

## 8 CONCLUSIONS

We raise a simple but important question: *Can we build an updatable learned index to fully replace the traditional disk-resident B+-tree?* To answer that, we explore state-of-the-art updatable learned indexes (expanding their in-memory implementations), and provide a comprehensive experimental evaluation and analysis. While each of them has their own strengths, we find none of them are competitive with the disk-resident B+-tree. Upon our evaluation, we present four crucial design choices. We believe our implementation and proposed design choices will be useful to researchers and practitioners in designing new and efficient disk-resident learned indexes.

## ACKNOWLEDGMENTS

This research is supported in part by ARC DP220101434 and DE230100366.